\documentclass[10pt,conference]{IEEEtran}
\AtBeginDocument{%
  }
\usepackage[linesnumbered,ruled,vlined]{algorithm2e}
\usepackage{booktabs}
\usepackage{amsmath,amssymb,amsfonts}
\usepackage{algpseudocode}
\usepackage{graphicx}
\usepackage{textcomp}
\usepackage[dvipsnames]{xcolor}
\usepackage{braket}
\usepackage{multirow}
\usepackage{xcolor,colortbl}
\usepackage{hyperref}
\usepackage{xspace}
\usepackage{subcaption}
\usepackage{enumitem}
\usepackage{pifont}
\usepackage{booktabs}
\usepackage[utf8]{inputenc}
\usepackage{tikz}
\newcommand{\bce}{\ell_{\mathrm{BCE}}}
\usepackage{balance}

\newif\ifcomments
\commentsfalse      

\newcommand{\acronym}[0]{QuFIG\xspace}

\begin{document}


\title{\acronym: GNN-Based Prediction of Quantum Fault Injection Vulnerabilities with Gate-Level Precision}

\author{
\IEEEauthorblockN{
Shihan Zhao\IEEEauthorrefmark{1}, 
Qiying Li\IEEEauthorrefmark{2}, 
Ben Dong\IEEEauthorrefmark{3}, 
Qian Wang\IEEEauthorrefmark{3}, 
Yuntao Liu\IEEEauthorrefmark{2}}
\IEEEauthorblockA{
\IEEEauthorrefmark{1}Department of Electrical \& Computer Engineering, Northeastern University, MA, USA \\
\IEEEauthorrefmark{2}Department of Electrical \& Computer Engineering, Lehigh University, PA, USA \\
\IEEEauthorrefmark{3}Department of Electrical Engineering, University of California, Merced, CA, USA \\
\IEEEauthorrefmark{1}zhao.shiha@northeastern.edu,
\IEEEauthorrefmark{2}\{qil420, yule24\}@lehigh.edu,
\IEEEauthorrefmark{3}\{cdong12, qianwang\}@ucmerced.edu}
}



\maketitle

\begin{abstract}
The growing scale and accessibility of quantum hardware exposed new reliability and security challenges in the quantum computing workflow, such as the run-time fault injection attacks in cloud-based quantum computing platforms. However, existing works fail to identify vulnerabilities with gate-level precision or adapt to run-time environments. In this work, we formulate gate-level fault analysis as a learning-guided prioritization problem under restricted fidelity budgets. 
The framework uses a circuit-DAG-based GNN backbone to predict the vulnerability score of each gate to each type of injected fault, defined as the impact of the gate-fault pair on circuit fidelity. The gate-fault pairs are then ranked by their vulnerability score.
Experiments on QASMbench and HamLib MaxCut show that \acronym\ recovers high-impact vulnerable gate-fault pairs with fewer inspections than random and depth-based heuristics. Our results show that \acronym can reduce the number of gates requiring inspection by 2.9--19.8\% while maintaining effective fault identification, allowing quantum circuit designers to identify vulnerabilities and apply targeted defenses more efficiently.
\end{abstract}
\section{Introduction}

The rapid evolution of quantum hardware has enabled increasingly large near-term workloads, including variational quantum algorithms (VQAs), quantum chemistry~\cite{peruzzo2014variational}, drug discovery~\cite{zhou2026quantum}, high-energy physics~\cite{andreassen2019junipr}, and quantum machine learning~\cite{biamonte2017quantum}. As quantum processors continue to scale, they can execute larger and more complex quantum circuits. However, because quantum hardware must operate in specialized cryogenic environments, it is typically accessed through cloud services rather than deployed locally. This reliance on remotely managed hardware introduces  security and trust challenges.

\begin{figure}[tb]
    \centering
    \includegraphics[width=0.9\linewidth]{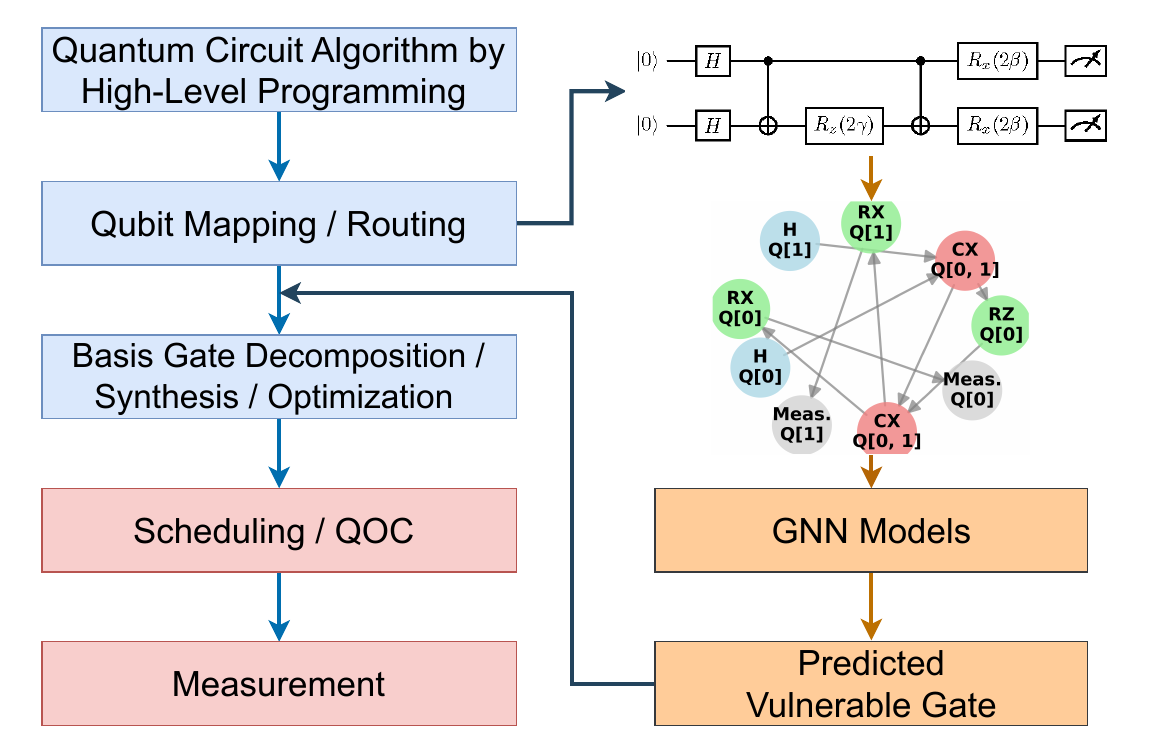}
    \caption{An overview of the quantum-circuit vulnerability prediction framework, which uses a GNN to identify potentially vulnerable gates.}
    \vspace{-5mm}
    \label{fig:overall}
\end{figure}

Cloud platforms such as IBM Quantum~\cite{ibmquantum}, Amazon Braket~\cite{gonzalez2021cloud}, and Microsoft Azure Quantum~\cite{prateek2023quantum} compile and execute remotely submitted circuits. As illustrated in Fig.~\ref{fig:overall}, compilation involves mapping, routing, decomposition, optimization, scheduling, and control before measurement. This managed workflow introduces security and trust risks when intermediate compilation or execution stages are compromised.\\
The quantum-computing workflow remains exposed to security and reliability risks across these stages. 
Existing studies mainly address higher-level vulnerabilities, including crosstalk-aware mapping and scheduling~\cite{deshpande2023design,ash2020analysis}, IP protection through obfuscated compilation~\cite{wang2025tetrislock, liu2025loq, zhang2025security}, and quantum trusted execution environments for workload isolation~\cite{trochatos2024dynamic}. 
However, these efforts largely overlook the gate level, where quantum programs are decomposed and synthesized into device-native gates before execution. 
A compromised decomposition or synthesis pipeline may inject or modify gates, causing fidelity degradation, energy-expectation shifts, or model-level mispredictions~\cite{chu2023qtrojan, john2025stealthy, dong2025qtime, zou2025conquer}. 
Thus, existing system-level defenses do not fully capture gate-level risks that directly threaten the physical integrity and measurement result.

A directed exhaustive gate-level fault analysis provides fine-grained vulnerability information by systematically injecting faults into individual gates according to a predefined fault model. The impact of each fault is then quantified by measuring the resulting fidelity loss or application-level output deviation.
Yet the complexity of this fine-grained analysis scales with the number of gates, fault types, circuits, and measurement shots, making it prohibitively expensive for large benchmarks. This suggests that gate-level quantum fault analysis should be performed selectively rather than exhaustively. 
Classical circuit testing follows a similar idea by using test points and observation points. Instead of monitoring every internal node, a small number of carefully chosen points are added to improve fault detection and diagnosis~\cite{Shi_Li_Khan_Wang_Wang_Huang_Xu_2022,Liang_Pinckney_Chai_Ren_Khailany_2023}.
Although quantum circuits do not support classical test points in the same physical sense, gate-level quantum fault analysis faces a closely related selection challenge: only a limited subset of gate-fault pairs contributes substantially to output degradation, while many faults remain weakly observable at the final measurement.

Based on this insight, we solve this problem by using graph learning-guided fault injection sensitivity analysis. 
For a compiled quantum circuit and a set of candidate NISQ fault types, we present the framework \acronym to rank gate-fault pairs by their predicted impact on circuit fidelity and output quality. 
The top-ranked candidates are then selected for fault injection or security auditing under a fixed evaluation budget.
In this way, the proposed method bridges fast backend-level simulation and precise gate-level fault analysis.
We summarize our contributions as follows:
\begin{itemize}[itemsep=1pt, parsep=0pt, topsep=1pt]
\item We formulate gate-level quantum fault analysis as a learning guided fault impact ranking problem, aiming to identify high-impact gate-fault pairs without exhaustive simulation.

\item We define a structured NISQ fault model covering stochastic decoherence faults, Pauli-type faults, and aggregate stochastic fault, and quantify fault impact using fidelity loss and application-level output deviation.

\item We develop a graph-based prediction framework \acronym that operates on compiled quantum circuit DAGs and ranks candidate gate-fault pairs for detailed simulation, mitigation, or security auditing.

  \item We evaluate \acronym under a budgeted vulnerable gate recovery setting on QASMbench and HamLib MaxCut circuits. At the top-16\% vulnerable-gate recovery target, \acronym reduces gate-inspection cost by 6.9\%--19.8\% relative to random selection and by 2.9\%--18.8\% relative to depth-based traversal. For the top-20\% vulnerable gates on QASMbench, \acronym further reduces the inspection cost by 9.9\% relative to random selection.

\end{itemize}
\section{Background} 
\subsection{NISQ Faults and Gate-Level Fault Injection}

As shown in Fig.~\ref{fig:overall}, a quantum circuit, similar to a classical circuit, is composed of a sequence of elementary logic operations. 
In quantum computing, each ideal gate can be represented as a unitary matrix acting on the quantum state~\cite{Duan_Liu_Fleming_Kompella_Xu_Ren_2025}. 
Given an input basis state $|x\rangle$, the circuit applies a unitary transformation $U$ and produces the output state
\begin{equation}
|\psi\rangle = U |x\rangle .
\end{equation}
For a compiled circuit consisting of multiple gate instances, the overall unitary can be decomposed as
\begin{equation}
U = U_n \cdots U_i \cdots U_2 U_1 ,
\end{equation}
Equivalently, using the density matrix representation
$\rho_0 = |\psi_0\rangle\langle\psi_0|$, the ideal output state can be written as
\begin{equation}
\rho_n
=
U_n \cdots U_1
\rho_0
U_1^\dagger \cdots U_n^\dagger .
\end{equation}

Gate-level fault injection simulates the effect of applying a fault to a specific gate instance. 
If a fault channel is injected after gate $U_i$, the intuitive state-vector form can be written as
\begin{equation}
|\psi_{\mathrm{faulty}}\rangle =
U_n \cdots \mathcal{E}_i U_i \cdots U_1 |\psi_0\rangle ,
\end{equation}
where $\mathcal{E}_i$ denotes the injected error at gate $U_i$. 
However, most NISQ noise sources, such as amplitude damping, phase damping, bit-flip, and depolarizing errors, are not ordinary unitary gates. 
They are non-unitary quantum channels and described using density matrices and Kraus operators:
\begin{equation}
\mathcal{E}_i(\rho)
=
\sum_k E_k \rho E_k^\dagger .
\end{equation}
Therefore, when the error channel is injected after $U_i$, the local update becomes
\begin{equation}
\tilde{\rho}_i
=
\mathcal{E}_i
\left(
U_i \rho_{i-1} U_i^\dagger
\right)
=
\sum_k
E_k
U_i \rho_{i-1} U_i^\dagger
E_k^\dagger .
\end{equation}
The remaining gates then continue to evolve this faulty state until the final measurement.


Because intermediate states cannot generally be observed without disturbing the computation, we quantify vulnerability by the fidelity or application-level deviation between ideal and faulty outputs. Exhaustive analysis is costly because the number of gate--fault candidates grows with circuit size and fault types, while an $n$-qubit output distribution spans up to $2^n$ basis states, motivating selective prioritization of high-impact candidates.
\subsection{GNN for Circuit Modeling}
GNNs learn structured representations on graph-structured data through iterative message passing, updating each node embedding by aggregating information from its neighbors:

\begin{equation}
\mathbf{h}_v^{(l+1)}
=
\sigma\!\left(
\mathbf{W}_1^{(l)}\mathbf{h}_v^{(l)}
+
\mathbf{W}_2^{(l)}
\frac{1}{|\mathcal{N}(v)|}
\sum_{u\in\mathcal{N}(v)}
\mathbf{h}_u^{(l)}
\right).
\label{eq:gnn-message-passing}
\end{equation}

where $\mathbf{h}_v^{(l)}$ is the hidden representation of gate node $v$ at layer $l$, 
$\mathcal{N}(v)$ is the set of neighboring gate nodes connected by qubit-wise dependencies, 
$\mathbf{W}_1^{(l)}$ and $\mathbf{W}_2^{(l)}$ are learnable transformations for self-node and neighborhood,
and $\sigma(\cdot)$ is a nonlinear activation. 
Each gate embedding incorporates local circuit context that may affect how a fault propagates.

In classical design automation, GNNs already handle tasks where circuit or program structure is naturally a graph: predict register-transfer-level timing, guide logic placement, optimize high-level synthesis scheduling, and manage resource allocations~\cite{wu2022high}. 

Quantum circuits map naturally onto DAGs, where nodes correspond to quantum gates and edges denote data or control dependencies. 
Recent work has used GNNs to improve noise-aware qubit mapping, gate scheduling, and layout optimization~\cite{liang2024graph,deshpande2023design, cheng2025epoc}. We extend this representation to a different task: ranking gate-level fault vulnerabilities in compiled quantum circuits. 
Rather than optimizing compilation cost, our model predicts how local perturbations or injected errors at individual gates affect the overall energy or fidelity of the circuit. 
By learning from simulated or empirical perturbation data, the GNN provides a vulnerability heatmap over the circuit’s DAG, highlighting fragile operations that may be exploited by noise or adversarial manipulation.

\section{Threat Model and Problem Formulation}
\label{sec:fault_model}

\subsection{Threat Model}

We consider a realistic cloud–based quantum computing environment in which victims develop quantum algorithms and rely on cloud quantum processors from providers such as IBM, Amazon Braket, and Google. 
We consider run-time error injection attacks. Specifically, the goal of the adversary is to maximally degrade the performance of the quantum circuit with minimal manipulations to avoid detection. Since the user cannot independently verify the security of cloud quantum computing environments, it is very important to identify “weak points” in the quantum circuit and apply targeted mitigation. The proposed vulnerability prediction method allows cloud quantum computer users to preemptively deploy defenses before deploying their workloads.

\subsection{Gate-Fault Pair}

We consider gate-level fault analysis on a compiled quantum circuit. 
Let $C$ denote a compiled circuit and let $G=\{g_1,\ldots,g_n\}$ be the set of gate instances in $C$. 
A candidate fault is defined as a gate-fault tuple
\[
f=(g,e,\eta),
\]
where $g\in G$ is the target gate instance, $e$ is the fault type, and $\eta$ denotes the fault strength, such as the error probability of a stochastic channel or the rotation offset of a coherent perturbation. 
This formulation focuses on transient gate-instance faults during circuit execution rather than permanent device defects. 
The objective is to evaluate how much each candidate fault changes the final computation and to prioritize the high-impact gate-fault pairs for detailed simulation, mitigation, or security auditing.

\subsection{Gate-Level NISQ Fault Classes}
\begin{table}[!htb]
\centering
\caption{Gate-level NISQ fault classes considered in this work.}
\label{tab:fault_classes}
\resizebox{0.95\linewidth}{!}{
\begin{tabular}{l|l|l}
\toprule
\textbf{Fault Class} & \textbf{Fault Type} & \textbf{Physical Interpretation} \\
\midrule
Stochastic Decoherence 
& \begin{tabular}[c]{@{}l@{}}Amplitude Damping\\Phase Damping\end{tabular}
& Relaxation and Dephasing \\
\midrule
Pauli-type fault 
& Bit flip 
& Logical State Perturbation \\
\midrule
Aggregate Stochastic Fault 
& Depolarizing 
& Effective Random Gate Error \\
\bottomrule
\end{tabular}
}
\vspace{-2mm}
\end{table}

We model gate-level faults using hardware-relevant NISQ error mechanisms. 
As summarized in Tab.~\ref{tab:fault_classes}, the considered fault channels are grouped into three classes: stochastic decoherence faults, Pauli-type faults, and aggregate stochastic faults. 
These fault classes cover the four noise channels used in our experiments, including amplitude damping, phase damping, bit flip, and depolarizing noise.

\textbf{Stochastic decoherence faults.}
These faults model probabilistic physical noise during circuit execution. 
Amplitude damping captures energy relaxation, where population in $|1\rangle$ may decay toward $|0\rangle$. 
Phase damping captures loss of phase coherence without directly changing computational-basis populations.

\textbf{Pauli-type faults.}
Pauli-type faults model discrete logical perturbations, such as bit-flip errors. 
They provide a simple gate-level abstraction of stochastic logical error manifestations and are useful for studying whether local logical perturbations propagate to the final measurement distribution.

\textbf{Aggregate stochastic faults.}
Depolarizing noise models effective randomization of the quantum state and is commonly used as an aggregate gate-level stochastic error model. 
It captures the combined effect of stochastic error sources when the detailed physical mechanism is not modeled separately.

For each valid gate--fault pair, we inject one of the considered noise channels at the target gate and execute the resulting circuit under the same simulation configuration as the fault-free circuit. We quantify the impact of each injection by comparing the faulty and ideal outputs using the corresponding fidelity or application-level metric. This procedure produces a gate--fault vulnerability score that serves as the ground-truth supervision for training and evaluation.

\subsection{Fault Activation and Observability}

Following the testing terminology of activation and observability, a fault at gate $g$ is activated when the input quantum state of $g$ has support on the subspace affected by the injected fault channel. 
For example, amplitude damping becomes observable only when the target qubit has non-negligible population in $\ket{1}$, while phase damping is most relevant when the qubit carries phase coherence. 
After activation, a fault is observable if its effect propagates to the final measurement distribution or to an application-level objective.

Given an ideal circuit $C_0$ and a faulty circuit $\tilde{C}_{g,e}$ where fault type $e$ is injected at gate $g$, we measure two forms of observability. 
First, we compute the distribution-level fidelity loss:
\[
\Delta_F(g,e)=1-\mathrm{Fid}(C_0,\tilde{C}_{g,e}).
\]
Second, for application-driven circuits such as MaxCut, we compute the normalized output deviation:
\[
\Delta_E(g,e)=
\frac{|E_{g,e}-E_0|}{|E_0|+\epsilon},
\]
where $E_0$ is the ideal objective value, $E_{g,e}$ is the objective value under the injected fault, and $\epsilon$ is a small constant to avoid numerical instability. 
The fidelity term captures output-distribution deviation, while the objective term captures application-level degradation.

\subsection{Fault Impact Score and High-Impact Labels}

We combine the two observability measures into a fault impact score:
\[
I(g,e)=\Delta_F(g,e)+\lambda \Delta_E(g,e),
\]
where $\lambda$ controls the relative importance of application-level deviation. 
For benchmarks without an application-specific objective, we set $\lambda=0$ and use fidelity loss as the impact metric.

Rather than relying on a fixed fidelity threshold, we define high-impact faults in a rank-based manner. 
For each circuit, we rank all valid gate-fault pairs according to $I(g,e)$ and label the top-$q\%$ as high-impact:
\[
Y[g,e]=
\begin{cases}
1, & \text{if } I(g,e) \in \mathrm{Top}\text{-}q\%,\\
0, & \text{otherwise},\\
-1, & \text{if fault type } e \text{ is not applicable to gate } g.
\end{cases}
\]
This rank-based definition avoids choosing an arbitrary global fidelity threshold and better matches the goal of budget-constrained fault-site prioritization. 
It also ensures that each circuit contributes a stable set of positive high-impact examples, even when absolute fidelity values differ across circuit families.

\subsection{Budget-Constrained Fault-Site Prioritization}

Exhaustive gate-level fault simulation evaluates all valid gate-fault pairs in a circuit. 
Let $\mathcal{F}$ denote the set of all valid candidate faults. 
Given a simulation or auditing budget $B$, our goal is to select a subset $\mathcal{S}_B \subseteq \mathcal{F}$ with $|\mathcal{S}_B|\le B$ that covers as many high-impact faults as possible:
\[
\max_{\mathcal{S}_B,\,|\mathcal{S}_B|\le B}
\sum_{(g,e)\in \mathcal{S}_B} I(g,e).
\]
Since computing $I(g,e)$ for every candidate pair requires exhaustive fault simulation, we train a graph-based predictor to approximate the ranking of gate-fault pairs from circuit structure. 
The learned ranking is then used to prioritize a small set of candidate faults for detailed simulation, mitigation, or security-oriented auditing.
\section{Learning-Guided Fault-Site Prioritization}
\label{sec:framework}

\begin{figure*}[t]
    \centering
    \includegraphics[width=0.9\linewidth]{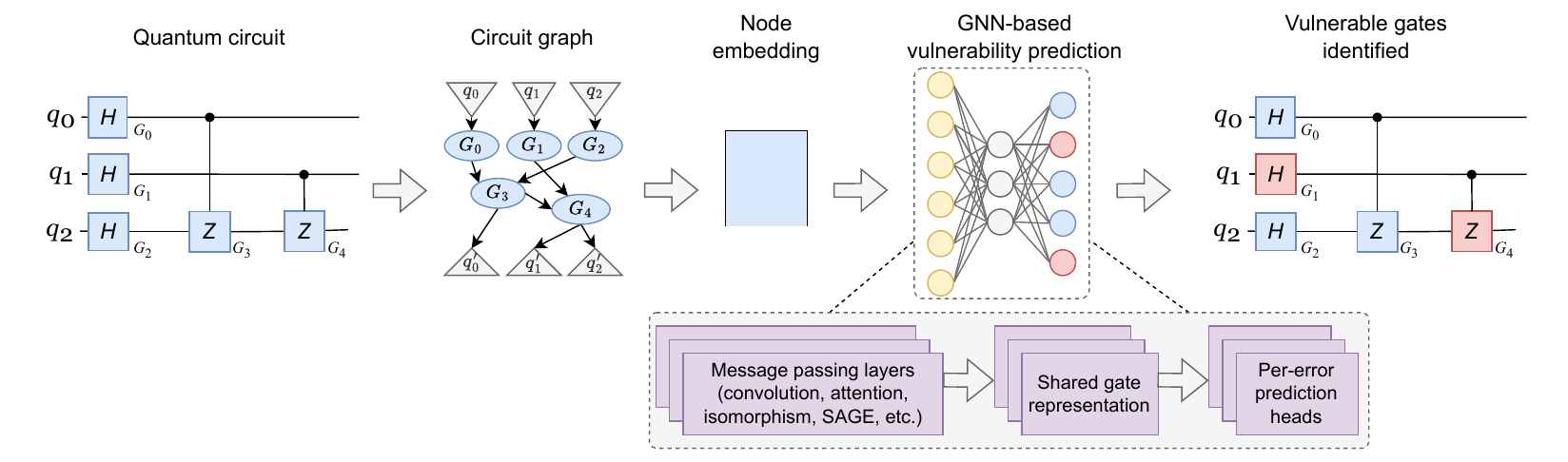}
    \caption{Overview of the proposed learning-guided fault-site prioritization framework. 
    A compiled quantum circuit is converted into a circuit DAG, where each gate node is encoded using gate-level structural features. 
    A GNN backbone estimates per-fault impact scores through fault-specific prediction heads, and the top-ranked gate-fault pairs are selected under a fixed budget for detailed fault simulation, mitigation, or security auditing.}
    \label{fig:workflow}
    \vspace{-2mm}
\end{figure*}

Based on the gate-level fault model defined in the previous section, we now describe the proposed learning-guided framework for fault-site prioritization. 
The goal is not to exhaustively simulate all candidate gate-fault pairs, but to predict and rank the subset of faults that are most likely to cause large output degradation. 
As shown in Fig.~\ref{fig:workflow} and Algorithm~\ref{alg:fault_prioritization}, the framework consists of five stages: circuit DAG construction, gate-level feature embedding, GNN-based fault-impact prediction, fault-aware training, and budget-constrained top-$k$ selection. 

\subsection{Circuit DAG Construction}

Given a compiled quantum circuit $C$, we convert it into a directed acyclic graph (DAG) $\mathcal{G}=(V,E)$, where each node $v\in V$ corresponds to a gate instance and each directed edge represents a qubit-wise execution dependency. 
If two gates are applied sequentially on the same qubit, a directed edge is added from the earlier gate to the later gate. 
This representation preserves the local propagation structure of the circuit and provides the basis for gate-level fault-impact prediction.

Each gate node is associated with a structured feature vector that captures its local circuit context. 
Tab.~\ref{tab:node_features} summarizes the gate-level features used in this work. 
These features describe both the functional role of each gate and its position in the compiled circuit.

\begin{table}[t]
  \centering
  \small
  \caption{Gate-level features used for circuit graph construction.}
  \label{tab:node_features}
  \resizebox{0.95\linewidth}{!}{
  \begin{tabular}{c|p{0.55\linewidth}|c}
    \toprule
    \textbf{Feature} & \centering\textbf{Description} & \textbf{Value / Range} \\
    \midrule
    \texttt{gate\_type} 
    & One-hot encoding of the quantum gate from a fixed basis set. 
    & 23-dim one-hot \\
    \texttt{num\_qubits} 
    & Number of qubits involved in the gate. 
    & $\{1,2\}$ \\
    \texttt{gate\_index} 
    & Topological index indicating the execution order. 
    & $[0,N_{\mathrm{gates}}{-}1]$ \\
    \texttt{qubit\_indices} 
    & Indices of the operated qubits, padded to a fixed length. 
    & 4-dim vector \\
    \texttt{num\_params} 
    & Whether the gate is parameterized. 
    & $\{0,1\}$ \\
    \bottomrule
  \end{tabular}}
  \vspace{-3mm}
\end{table}

\subsection{Gate Embedding and GNN Backbones}

Let $X\in\mathbb{R}^{|V|\times d}$ denote the node feature matrix of the circuit DAG, where each row corresponds to the feature vector of one gate instance. 
A learnable embedding layer first maps the raw gate features into a hidden representation. 
The embedded circuit graph is then processed by a message-passing GNN backbone following the general formulation in Eq.~\eqref{eq:gnn-message-passing}, so that each gate representation incorporates dependency information from neighboring gates in the circuit DAG.

Our framework is backbone-agnostic and does not introduce a new message-passing operator. 
Instead, we instantiate the predictor with standard GNN backbones, including graph convolutional network (GCN), graph attention network (GAT), graph isomorphism network (GIN), and graph sample and aggregate (GraphSAGE), and evaluate which inductive bias is most suitable for quantum fault-site prioritization.

\subsection{Per-Fault Impact Prediction}

Each gate can be affected differently by different NISQ fault types. 
Therefore, instead of predicting a single vulnerability label for each gate, the model predicts a fault-specific score for each valid gate-fault pair. 
Let $K$ denote the number of fault types. 
For each gate $v$ and fault type $k$, the model outputs a logit $z_{v,k}$ through a fault-specific prediction head:
\begin{equation}
z_{v,k} = \mathrm{Head}_k(\mathbf{h}_v),
\end{equation}
and converts it into a high-impact probability:
\begin{equation}
p_{v,k} = \sigma(z_{v,k}).
\end{equation}

A binary mask $m_{v,k}$ is used to indicate whether fault type $k$ is applicable to gate $v$. 
Invalid gate-fault pairs are excluded from both training and evaluation. 
The resulting score matrix $P\in[0,1]^{|V|\times K}$ provides a ranked estimate of which gate-fault pairs are most likely to cause large output degradation.



\subsection{Budget-Constrained Top-k Selection}

After inference, the predicted scores are used to rank all valid gate-fault pairs in the circuit. 
Given a simulation, mitigation, or auditing budget $B$, we select the top-$B$ candidates:
\begin{equation}
\mathcal{S}_B
=
\mathrm{TopB}_{(v,k)}
\left(
p_{v,k}
\right),
\quad
\text{s.t. } m_{v,k}=1.
\end{equation}
The selected set $\mathcal{S}_B$ contains the gate-fault pairs that are predicted to have the largest impact on the final computation. 
These candidates are then passed to detailed fault simulation, mitigation, or security auditing. 
This step reduces the number of gate-fault pairs that need to be exhaustively simulated while preserving the high-impact faults most likely to dominate output degradation.

\begin{algorithm}[!t]
\DontPrintSemicolon
\SetAlgoLined
\caption{Learning-Guided Quantum Fault-Site Prioritization}
\label{alg:fault_prioritization}

\KwIn{Compiled circuit $C$, candidate fault types $\mathcal{E}$, predictor $f_\theta$, budget $B$}
\KwOut{Prioritized fault set $\mathcal{S}_B$}

\tcp{Circuit graph construction}
$\mathcal{G}=(V,E) \gets \mathrm{DAG}(C)$\;
$X \gets \mathrm{ExtractGateFeatures}(\mathcal{G})$\;

\tcp{Fault-impact prediction}
\ForEach{$v \in V$}{
    \ForEach{$k \in \mathcal{E}$}{
        \If{$(v,k)$ is valid}{
            $p_{v,k} \gets f_\theta(\mathcal{G}, X, v, k)$\;
        }
    }
}

\tcp{Budget-constrained selection}
$\mathcal{S}_B \gets \mathrm{TopB}_{(v,k)}(p_{v,k})$\;

\KwRet{$\mathcal{S}_B$}
\end{algorithm}

To encourage well-calibrated prediction confidence, we use a hybrid loss combining binary cross-entropy (BCE) and mean absolute error (MAE). The BCE term provides the primary classification signal, encouraging the model to correctly identify whether a gate is vulnerable under a specific noise channel. BCE enforces discriminative predictions and is effective for separating vulnerable gates from non-vulnerable ones, while MAE complements BCE by penalizing low-confidence predictions.
\begin{align}
\mathcal{L} 
&= \mathcal{L}_{\mathrm{BCE}} + \mathcal{L}_{\mathrm{MAE}}, \\
\mathcal{L}_{\mathrm{BCE}} 
&= \frac{1}{K} \sum_{k=1}^{K} 
\frac{1}{\sum_v m_{v,k}}
\sum_{v} m_{v,k}\,
\bce(z_{v,k}, y_{v,k}; \alpha_k), \\
\mathcal{L}_{\mathrm{MAE}} 
&= \frac{1}{K} \sum_{k=1}^{K} 
\frac{1}{\sum_v m_{v,k}}
\sum_{v} m_{v,k}\,
\lvert p_{v,k} - y_{v,k} \rvert .
\end{align}
\noindent
where $m_{v,k}$ masks out inapplicable error types and 
$\alpha_k = N_k^{\mathrm{neg}} / N_k^{\mathrm{pos}}$ compensates for class imbalance.
For each gate $v$ and error type $k$, the model outputs a logit
$z_{v,k}$ and a corresponding probability $p_{v,k}=\sigma(z_{v,k})$, which is compared against the binary ground-truth label $y_{v,k}\in\{0,1\}$. 
The MAE loss penalizes the absolute deviation $|p_{v,k}-y_{v,k}|$, encouraging the predictor to assign probabilities close to $1$ for truly vulnerable gates ($y_{v,k}=1$) and close to $0$ for non-vulnerable ones ($y_{v,k}=0$).
This explicitly suppresses low-confidence or borderline predictions (e.g., $p_{v,k}\approx 0.5$), even when such predictions might still yield a correct binary classification under the standard $0.5$ threshold. 
Comparing with BCE focuses on classification correctness, the MAE term acts directly on the predicted probabilities. By integrating BCE and MAE, the model benefits from both strong classification gradients and improved confidence calibration.

\section{Experiments}
\subsection{Experiment Setup}

\textbf{Circuit datasets.} We evaluate our method on two suites of quantum circuit datasets, QASMbench~\cite{li2022qasmbench} and HamLib MaxCut~\cite{sawaya2023hamlib}, which represent different sources of circuit diversity.   
For QASMbench, we used circuits from both the small-scale and medium-scale subsets. 
Since individual QASMbench subsets may contain limited structural variations, we concatenate the small and medium circuits to augment the dataset and expose the model to a broader range of circuit sizes, gate compositions, and dependency structures. 
For HamLib MaxCut, 12 base problem graphs are expanded by varying the cost-layer type, mixer-layer type, and QAOA depth. 
This gives 2304 candidate circuits, among which 2120 valid compiled circuits are used in our experiments. The details of the dataset is shown in Tab.~\ref{tab:dataset_scale} and Tab.~\ref{tab:maxcut_structure} 

\begin{table}[!htbp]
\centering
\caption{Dataset scale used in our experiments.}
\label{tab:dataset_scale}
\begin{tabular}{lcccc}
\toprule
Dataset & Orig. Circuits & Aug. Circuits & Gate Range  \\
\midrule
QASMbench~\cite{li2022qasmbench} & 35 & 2907 & 4--2053  \\
Hamlib-MaxCut~\cite{sawaya2023hamlib} & 2120 & 0 & 7--342  \\
\bottomrule
\end{tabular}
\end{table}

\textbf{Ground-truth generation.}
For each circuit, we first compile it into a directed acyclic graph (DAG) where each node denotes a quantum gate, and each edge captures the qubit-level dependency between gates, then perform gate-level fault injection using IBM Qiskit simulation~\cite{qiskit2024}. 
For each valid gate--error pair, we inject one of the considered noise channels, mainly amplitude damping, phase damping, bit flip, and depolarizing noise, and compare the perturbed output with the ideal output. 
The resulting fidelity degradation is used to determine whether a gate is vulnerable under a specific error type. 
This process produces multi-label gate-level supervision, where each gate can be associated with multiple vulnerability labels corresponding to different noise channels.
\begin{table}[!htbp]
\centering
\caption{HamLib MaxCut circuit generation factors. Circuits are reported as valid/candidate counts.}
\label{tab:maxcut_structure}
\begin{tabular}{ccccc}
\toprule
Base Graphs & Cost Types & Mixer Types & Depths & Valid Circuits \\
\midrule
12 & 6 & 8 & 1, 2, 3, 5 & 2120 \\
\bottomrule
\end{tabular}
\end{table}

\textbf{Model setting.}
GNN models are implemented with PyTorch Geometric~\cite{fey2019fast}. 
We use GraphSAGE, denoted as SAGE in our implementation, as the main backbone model because its neighborhood aggregation mechanism naturally matches the local dependency structure of quantum circuit DAGs. 
Each model is empirically configured with four GNN layers and a hidden dimension of 128. 
For each gate node, the model predicts vulnerability scores over different error types. 
The output logits are optimized using the hybrid BCE--MAE loss described above, where BCE provides the primary classification signal and MAE encourages calibrated prediction confidence. Models are trained using the Adam optimizer for 200 epochs with a learning rate of $1\times10^{-4}$. 

\textbf{Baselines.}
We compare our vulnerability-guided insertion strategy with several heuristic baselines. 
The first baseline is random insertion, where gates are selected uniformly at random under the same budget. 
We also consider depth-based insertion strategies, including selecting gates from the beginning of the circuit depth order, from the end of the circuit depth order, and from the middle-depth region outward. 
These baselines represent simple but intuitive circuit-structure heuristics and allow us to evaluate whether the learned SAGE predictor provides additional benefit beyond depth-based gate prioritization.

For the learning-based method, datasets are randomly split into 80\% training, 10\% validation, and 10\% test sets. 
Since the number of available quantum circuits is limited, we further adopt a 10-fold cross-validation strategy to obtain a more reliable estimate of model performance. 
To avoid structural information leakage, the split is performed in a group-aware manner. 
Specifically, when a certain MaxCut circuit structure or a QASMbench circuit is assigned to the test set, the corresponding circuit itself and all concatenated circuits using it as a structural pattern are excluded from the training set. 

\begin{figure}[t]
    \centering
    \includegraphics[width=0.95\linewidth]{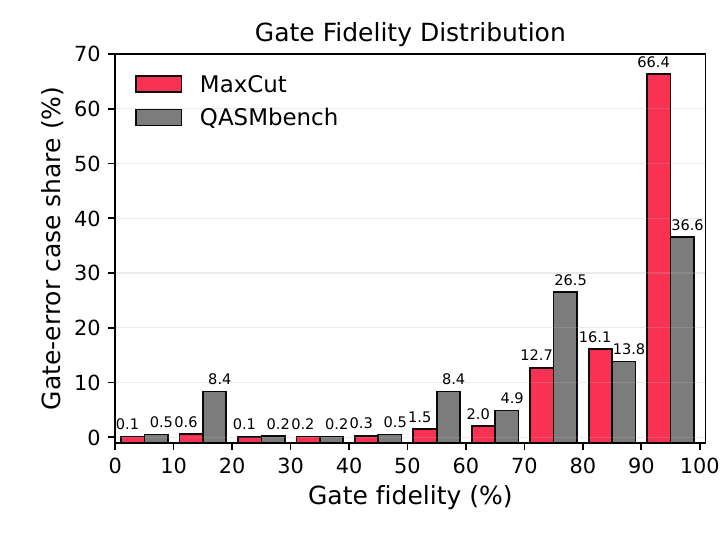}
    \caption{Gate-level fidelity distribution under single-gate fault injection. The $x$-axis groups injected gate--error cases by the output fidelity between the faulty and ideal circuits, and the $y$-axis reports the percentage of cases in each bin. Lower fidelity indicates stronger output degradation. For example, only 0.1\% and 0.5\% cases reduce the output fidelity to lower than 10\%, showing that severe faults are concentrated in a small subset.} 
    \vspace{-3mm}
    \label{fig:gate_fidelity_distribution}
\end{figure}
\subsection{Quantum Circuit Error Fidelity Analysis}
Fig.~\ref{fig:gate_fidelity_distribution} shows the distribution of gate-level output fidelity under high-strength fault injection. 
The results reveal that fault sensitivity is highly non-uniform across both gates and datasets, rather than being evenly distributed throughout the circuit. 
As expected, only a small subset of gates causes substantial output degradation after noise injection, whereas most gates remain relatively stable and exhibit only minor fidelity fluctuations even under strong perturbations. 
The gates that can truly overturn the circuit outcome are rare, while the majority of gates are robust.

\begin{figure}[!htbp]
    \centering
    \includegraphics[width=0.95\linewidth]{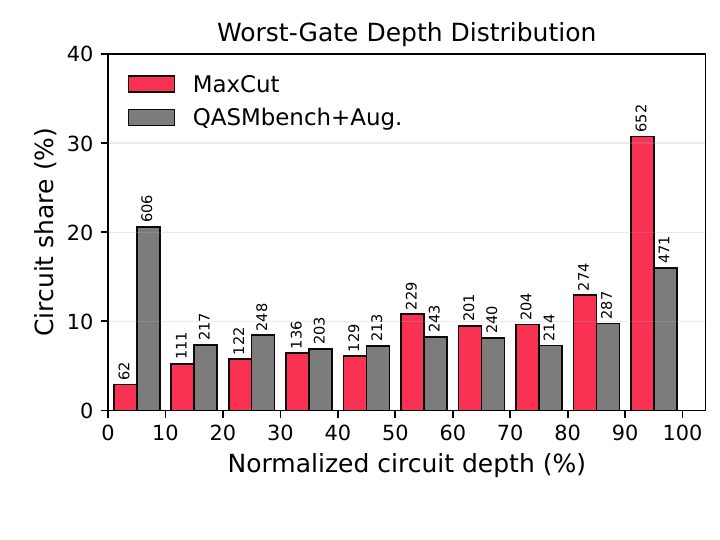}
    \caption{Position distribution of the most vulnerable gates. 
    For each circuit,  The $x$-axis shows 10\% position bins which compute most vulnerable gate normalized position as $\text{gate index}/\text{total gates}$. 
    The $y$-axis reports the the percentage of circuits whose most vulnerable gate falls into each bin. 
    Numbers above bars indicate circuit counts. 
    The dataset-dependent distribution shows that simple position-based traversal is insufficient for reliable vulnerable-gate detection.}

    \vspace{-3mm}
    \label{fig:worst_gate_depth_distribution}
\end{figure}
To further examine whether vulnerable gates are associated with specific depth regions, we identify, for each circuit, the top 10\% most vulnerable gates ranked by gate impact and group their normalized depth positions into 10\% bins. 
As shown in Fig.~\ref{fig:worst_gate_depth_distribution}, the vulnerable gates do not follow a consistent depth-dependent pattern across circuits. 
Some circuit families exhibit more vulnerable gates in later stages, whereas others show higher noise sensitivity in earlier or middle regions. 
This indicates that circuit depth alone does not directly determine gate vulnerability. 
Instead, the location of vulnerable gates varies substantially with circuit structure, suggesting that vulnerability is governed by the interaction between gate position, local dependency context, and overall circuit organization.
\subsection{GNN Backbone Analysis}
\begin{table}[!htp]
\centering
\small
\caption{Backbone comparison on QASMbench and MaxCut. MAE measures score calibration, while AUPRC measures ranking quality under class imbalance.}
\label{tab:gnn_backbone_comparison}
\begin{tabular}{lcccc}
\toprule
\multirow{2}{*}{GNN} & \multicolumn{2}{c}{QASMbench} & \multicolumn{2}{c}{MaxCut} \\
\cmidrule(lr){2-3} \cmidrule(lr){4-5}
 & MAE $\downarrow$ & AUPRC $\uparrow$ 
 & MAE $\downarrow$ & AUPRC $\uparrow$ \\
\midrule
GCN  & 0.225 & 0.191 & \underline{0.223} & \underline{0.273} \\
GAT  & \textbf{0.215} & 0.200 & 0.228 & 0.242 \\
GIN  & 0.227 & \underline{0.213} & 0.227 & 0.267 \\
SAGE & \underline{0.218} & \textbf{0.220} & \textbf{0.216} & \textbf{0.283} \\
\bottomrule
\end{tabular}
\end{table}
Tab.~\ref{tab:gnn_backbone_comparison} compares different GNN backbones on QASMbench and MaxCut using MAE and AUPRC. 
MAE evaluates whether the predicted vulnerability scores are well calibrated, while AUPRC evaluates whether the model can rank vulnerable gate-fault pairs effectively under class imbalance. 
On QASMbench, SAGE achieves the lowest MAE, indicating slightly better score calibration, whereas SAGE obtains the highest AUPRC, suggesting better ranking quality for vulnerable gates. 
On MaxCut, SAGE achieves the best performance on both MAE and AUPRC, showing that its neighborhood aggregation mechanism is well-suited to capturing the local dependency structure of quantum circuit DAGs. 
Overall, SAGE provides the most consistent performance across the two datasets.
Therefore, we use SAGE as the main backbone.

\subsection{Vulnerablity Point Detection}
\begin{figure}[!htbp]
    \centering
    \includegraphics[width=0.95\linewidth]{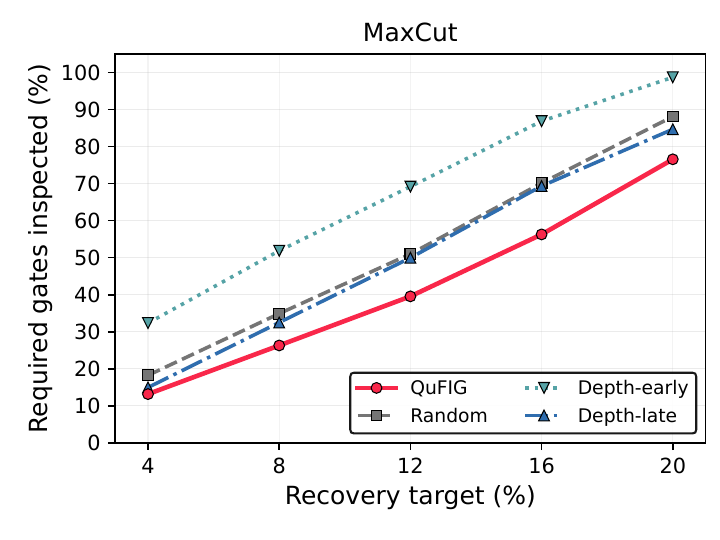}
    \caption{Target-recovery cost on MaxCut. 
SAGE achieves the lowest inspection cost compared with random selection, depth-early selection that scans gates from the beginning to the end according to gate depth index, and depth-late selection that scans gates in the reverse depth order. }
    \vspace{-3mm}
    \label{fig:target_point_maxcut}
\end{figure}
\begin{figure}[!htbp]
    \centering
    \includegraphics[width=0.95\linewidth]{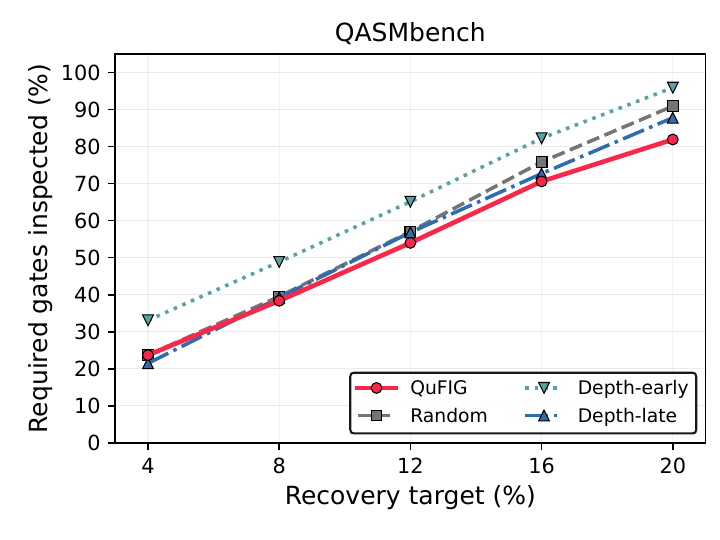}
    \caption{Target-recovery cost on QASMbench. \acronym achieves the lowest inspection cost compared with random selection, depth-early selection and depth-late selection.}
    \vspace{-3mm}
    \label{fig:target_point_qasmbench}
\end{figure}

\subsection{Target-detection Cost Evaluation}
Fig.~\ref{fig:target_point_maxcut} and Fig.~\ref{fig:target_point_qasmbench} report the point-insertion cost required to identify the top-$K\%$ most vulnerable gates. 
The $x$-axis denotes the target vulnerable-gate set, ranging from the top 4\% to the top 20\% gates ranked by ground-truth vulnerability, and the $y$-axis denotes the required gate-inspection budget. 
Lower values indicate better prioritization, since fewer gates need to be inspected to recover the same set of highly vulnerable gates.

On the MaxCut dataset, the \acronym ranking consistently requires fewer inspected gates than random selection and simple heuristic traversals. 
For example, to identify the top 16\% vulnerable gates, the \acronym needs to inspect only 56.3\% of gates, while random selection requires 70.1\% and depth-late traversal requires 69.3\%. 
This shows that the learned model captures vulnerability patterns that cannot be explained by simply scanning gates according to circuit depth or backend error rates.

On QASMbench, the \acronym also reduces the required inspection cost compared with random selection, especially when the target vulnerable set becomes larger. 
To identify the top 16\% vulnerable gates, the \acronym inspects 70.6\% of gates, compared with 75.8\% for random selection and 72.7\% for depth-late traversal. 
For the top 20\% vulnerable gates, the \acronym requires 81.9\% of gates, while random selection requires 90.9\%. 
These results indicate that the \acronym provides a more effective ranking of vulnerable gates than uninformed testing, although the improvement is smaller than on MaxCut due to the stronger depth-related structure in QASMbench. Overall,  Instead of exhaustively inspecting all gates, the \acronym ranks gates by predicted vulnerability and recovers a large fraction of the worst gates with fewer inspections, particularly on structurally diverse MaxCut circuits.

\section{Conclusion}
In this work, we presented \acronym for budget-constrained gate vulnerability prioritization in quantum circuits. 
We construct gate-level fault-injection datasets from QASMbench and HamLib MaxCut circuits, covering diverse gate types, circuit depths, dependency structures, and NISQ fault channels. 
The results show that learned vulnerability scores can prioritize high-impact gates and reduce the number of detailed gate inspections needed to reach a target recovery level.
Compared with random and simple structural traversal strategies, \acronym ranking more effectively identifies gates that dominate circuit-level fidelity loss and application-level degradation.
By identifying the gates that dominate fidelity loss before deployment, \acronym lets users of untrusted cloud quantum platforms concentrate fault simulation, mitigation, and security auditing where it matters most.

\section*{Acknowledgment}
This work is supported by the National Science Foundation under Awards 2530705, 2554928, and 2554929.

\balance
\bibliographystyle{IEEEtran}
\bibliography{reference}

\end{document}